\documentstyle[pre,aps]{revtex}

   \font\tenmsb=msbm10 scaled\magstep 1
   \font\sevenmsb=msbm7 scaled \magstep 1
   \font\faivemsb=msbm5 scaled \magstep 1
\newfam\msbfam
      \textfont\msbfam=\tenmsb
      \scriptfont\msbfam=\sevenmsb
      \scriptscriptfont\msbfam=\faivemsb
\def\Bbb#1{{\fam\msbfam #1}}
\font\tengothic=eufm10 scaled\magstep 1
\font\sevengothic=eufm7 scaled\magstep 1
\newfam\gothicfam
      \textfont\gothicfam=\tengothic
      \scriptfont\gothicfam=\sevengothic

\newcommand{\ra}{\rightarrow}
\newcommand{\be}{\begin{equation}}
\newcommand{\ee}{\end{equation}}
\newcommand{\dt}{\delta}
\newcommand{\lm}{\lambda}
\newcommand{\Lm}{\Lambda}
\newcommand{\al}{\alpha}
\newcommand{\bt}{\beta}
\newcommand{\ga}{\gamma}
\newcommand{\kp}{\kappa}
\newcommand{\ep}{\varepsilon}

\begin{document}

\draft
\title{Temporal Dynamics in Perturbation Theory} 
\author{V. I. Yukalov and E. P. Yukalova}
\address{Department of Mathematics and Statistics\\
Queen's University, Kingston, Ontario K7L 3N6, Canada\\
and \\
Bogolubov Laboratory of Theoretical Physics \\
Joint Institute for Nuclear Research, Dubna 141980, Russia} 
\maketitle

\vspace{1cm}

\begin{abstract}

Perturbation theory can be reformulated as dynamical theory. Then a 
sequence of perturbative approximations is bijective to a trajectory of 
dynamical system with discrete time, called the approximation cascade. 
Here we concentrate our attention on the stability conditions permitting 
to control the convergence of approximation sequences. We show that several 
types of mapping multipliers and Lyapunov exponents can be introduced and, 
respectively, several types of conditions controlling local stability 
can be formulated. The ideas are illustrated by calculating the energy 
levels of an anharmonic oscillator.
\end{abstract}

\section{Introduction}

Perturbation theory as applied to realistic physical problems practically 
always yields divergent sequences. There is a number of methods of finding 
effective limits of such sequences. The most known are the methods of Pad\'e 
and of Borel summation. Constantly, new techniques appear. We are not aiming 
to give here a review of those. But we think it would be worth mentioning some
of the recently developed interesting methods. Among them are some variants of
the Pad\'e method [1,2] and the converse Cesaro method [3]. An analogy of the 
Kolmogorov perturbation theory in classical mechanics has been constructed for
self--adjoint operators [4], which reorganizes the asymptotic series in the 
usual Rayleigh--Schr\"odinger perturbation theory to a generalized asymptotic 
series. The accuracy of perturbation theory is known to improve when a 
perturbative expansion is accompanied by a variational procedure [5-7]. The 
latter often gives good results already at the first step of calculations, as 
in the method of potential envelopes [8,9] or when the zero approximation is 
very close to the considered problem [10]. The variational principle improves 
the accuracy of the quasiclassical and Langer methods [11] and of different 
sequence transformations [12]. The critical behaviour of several spin and 
vertex models has been studied by combining a variational series expansion 
and the coherent anomaly method [13-17].

The common drawback of practically all perturbation techniques used for 
physical problems is the absence of a general method for checking the 
convergence of perturbation sequences. This is because an explicit expression 
for the perturbative approximation of arbitrary order is, except a few trivial
cases, never available. For some simple problems one can obtain exact 
numerical solutions. Then the accuracy of perturbation theory can be checked 
by a direct comparison of its results against the known exact answers. If, 
when increasing the approximation order, one gets better accuracy, then it 
is common to say that the procedure is convergent. Strictly speaking, this 
terminology is not correct, since the classical notion of convergence 
presupposes that the sequence of approximations has an exact solution as its 
limit when the approximation order tends to infinity. However, in actual 
calculations one always has a finite number of approximations, though the 
approximation order may be quite high. There is a number of examples when 
the accuracy of a calculational scheme improves till a fixed approximation 
order but after it worsens resulting in a divergent sequence. Such behaviour 
is, indeed, a common feature of asymptotic series. In this situation it is 
more correct to say that the sequence of perturbation theory is 
semiconvergent.

Note that here and in what follows we use the term "perturbation theory" in 
its most general meaning: perturbation theory is a regular procedure 
prescribing a general algorithm for defining a sequence of approximations of 
arbitrary order. The term "general algorithm" means that there is a general 
rule for calculating any approximation, although in practice it may happen 
that because of technical difficulties, one is able to find just a few initial
approximations. The algorithm of perturbation theory may include, in addition 
to a basic expansion procedure, any of various resummation or renormalization 
tricks. In this sense, the so--called nonperturbative approaches are nothing 
but particular variants of perturbation theory, supplied with additional 
conditions.

The main question for any variant of perturbation theory in the case of a 
complicated problem is how to control the convergence if neither an explicit 
form of high--order approximations nor the exact solution are available? To 
overcome this difficulty, an idea has been advanced [18] that the perturbation
algorithm can be supplemented by a set of functions controlling the 
convergence of the approximation sequence. These functions because of their 
role, can be called the control, or governing, functions. Perturbation theory 
employing control functions has been first published in ref.[19] and used for 
describing anharmonic crystals [20-25]. In the cited papers, control functions
were defined by a minimal--difference condition. A variational approach, 
called the minimal--sensitivity condition [5], has also been applied to 
several anharmonic models [26-34].The choice of conditions for control 
functions has been heuristic.

To justify the option of conditions defining control functions, it was shown 
[35-39] that perturbation theory can be formulated as renormalization--group
theory. Then control functions are to be defined from a fixed--point 
condition, whose particular variants yield either the minimal--difference or 
minimal--sensitivity conditions. As far as a renormalization group can be 
considered as a kind of dynamical system, it was natural to reformulate 
perturbation theory to the language of dynamical theory [40-42]. This 
reformulation not only makes the theory more logical but also permits to 
define stability conditions related to the problem of convergence. In our 
previous papers we did not pay enough attention to these stability problems. 
The purpose of the present publication is to compensate this defficiency. We 
proffer a detailed analysis of the stability conditions showing that, really, 
there are several types of them, each with a different meaning. These 
conditions allow to understand intimate features of perturbation sequences 
and, therefore, to control their properties. Perturbation theory, whose 
convergence is controlled by control functions and which is supplemented  
by the stability conditions allowing for a thorough control of the properties 
of perturbation sequences, can be called the {\it controlled perturbation 
theory}.

\section{Survey of approach}   

Before turning to stability conditions, we need to give a brief survey of 
the method whose stability is to be analysed.

Assume that we are trying to solve a problem whose solution is a function 
$\;f(g)\;$, such that $\;f:\;\Bbb{D}\ra\Bbb{R}\;$, of a variable 
$\;g\in\Bbb{D}\subset\Bbb{R}\;$. The first thing we have to do is to 
introduce {\it control functions}. For simplicity, we shall speak here about 
one set of such functions. The generalization to the case of several sets is 
straightforward [42]. Incorporate into the perturbation algorithm a parameter
$\;u\in\Bbb{R}\;$ whose value is yet undefined. Then we get a sequence
$\;\{ F_k(g,u)\}_{k=0}^\infty\;$ of functions $\;F_k:\;\Bbb{D} \times 
\Bbb{R}\ra\Bbb{R}\;$. In each term $\; F_k(g,u)\;$, we substitute for the 
parameter $\;u\;$ a function $\;u_k(g)\;$, such that $\;u_k:\;\Bbb{D} \ra 
\Bbb{R}\;$ and $\;k\in\Bbb{Z}_+ \equiv \{ 0,1,2,\ldots \}\;$. This results in 
a function $\;f_k:\;\Bbb{D}\ra\Bbb{R}\;$ which is
\be
f_k(g)\equiv F_k(g,u_k(g)) .
\ee
The role of the control functions $\;u_k(g)\;$ is to govern the convergence 
of the sequence $\;\{ f_k(g)\}_{k=0}^\infty\;$ to the limit
\be
\lim_{k\ra\infty}f_k(g) = f(g) .
\ee
As is clear, there can be infinite number of different types of functions 
satisfying (2). This means that (2) defines a class of equivalence
$${\cal U} \equiv \{u_k(g)|\; k\in\Bbb{Z}_+\} . $$
The latter can be explicitly found only if the general form of (1), as 
$\;k\ra\infty\;$, is known. Of course, this can be realised only for a few 
simple zero--and one--dimensional models [43-45]. Usually, the explicit form 
of (1) for arbitrary $\;k\ra\infty\;$ is not known.

The second step is to narrow the class of equivalence $\;\cal U\;$ by those 
control functions that satisfy an evolution equation defining a dynamical 
system in discrete time, called the {\it approximation cascade}. This task is 
a kind of an inverse problem in the optimal control theory [46,47]. The direct
problem in the latter is when an evolution equation is given and one has to 
find control functions minimizing a performance index. In our case, we need 
to find an evolution equation itself. To this end, introduce the coupling 
function $\;g_k(f)\;$, such that $\;g_k:\;\Bbb{R}\ra\Bbb{D}\;$, given by the 
equation
\be
F_0(g_k(f),u_k(g_k(f)))=f .
\ee
As is clear, the function $\;g^{-1}_k:\;\Bbb{D}\ra\Bbb{R}\;$ defined as 
\be
g_k^{-1}(g) \equiv F_0(g,u_k(g))
\ee
is inverse with respect to $\;g_k(f)\;$, since
\be
g_k^{-1}(g_k(f)) =f, \qquad g_k(g_k^{-1}(g)) = g .
\ee
Introduce the function
\be
y_k(f)\equiv f_k(g_k(f))
\ee
realizing an endomorphism $\;y_k:\;\Bbb{R}\ra\Bbb{R}\;$ of the measurable 
space $\;\Bbb{R}\;$. The function (1) can be recovered from (6) by the 
transformation
\be
f_k(g) =y_k(g_k^{-1}(g)) .
\ee
The sequence $\;\{ y_k(f)\}_{k=0}^\infty\;$, by construction, is bijective to
$\;\{ f_k(g)\}_{k=0}^\infty\;$.

Let us require that the endomorphism $\;y_k\;$ possess the semigroup
properties
\be
y_k\cdot y_p =y_{k+p} , \qquad y_0 =1 .
\ee
This is equivalent to the relation
\be
y_{k+p}(f) =y_k(y_p(f)) ,
\ee
having the meaning of an evolution equation with the initial condition 
$\;y_0(f) =f\;$. Equations analoguos to (9) can be met in various physical 
problems where they are often called the self--similar relations [48]. The 
requirement (8) narrows the class of equivalence $\;\cal U\;$ to those control
functions that provide the validity of the evolution equation (9).

The semigroup
\be
\Bbb{Y} \equiv \{ y_k\}:\; \Bbb{Z}_+\times\Bbb{R}\ra\Bbb{R}
\ee
of the endomorphisms $\;y_k\;$ defines a dynamical system with the discrete 
time $\;k\in\Bbb{Z}_+\;$. In the dynamical theory this is called a 
semicascade. In our case, the latter is related to the sequence of 
approximations (7) because of which we call (10) the approximation cascade. 
The cascade trajectory $\;\{ y_k(f)\}_{k=0}^\infty\;$ is bijective to the 
approximation sequence $\;\{ f_k(g)\}_{k=0}^\infty\;$. The existence of the 
limit (2), which, according to (7), can be written as
\be
\lim_{k\ra\infty}f_k(g) =\lim_{k\ra\infty}y_k(g_k^{-1}(g)) = f(g) ,
\ee
is equivalent to the existence of an attracting fixed point of the 
approximation cascade,
\be
\lim_{k\ra\infty}y_k(f) =\lim_{k\ra\infty}f_k(g_k(f)) =y^*(f) .
\ee

At the third step, we embed the approximation cascade (10) into an 
{\it approximation flow}. This is done as follows. Instead of the discrete 
variable $\;k\in\Bbb{Z}_+\;$ consider a continuous variable $\;t\in\Bbb{R}_+ 
\equiv [0,\infty )\;$. Introduce an endomorphism $\;y(t,\cdot ):\; 
\Bbb{R}\ra\Bbb{R}\;$ of the measurable space$\;\Bbb{R}\;$, satisfying the 
semigroup properties (8) for each $\;t\in\Bbb{R}_+\;$. Then in the place of 
(9) we have
\be
y(t+t',f) = y(t,y(t',f)) .
\ee
Require that the endomorphism $\;y(t,\cdot )\;$ would satisfy the conditions

$$ y(k,f) = y_k(f) \qquad (k\in\Bbb{Z}_+) , $$
\be
\lim_{t\ra\infty}y(t,f) =y^*(f) .
\ee 
The semigroup
\be
\stackrel{\sim}{\Bbb{Y}} \equiv \{ y(t,\cdot )\}:\; \Bbb{R}_+\times\Bbb{R} 
\ra \Bbb{R}
\ee
of the endomorphisms $\;y(t,\cdot )\;$ is a dynamical system called semiflow. 
We name (15) the approximation flow. By construction, the semigroup (10) is a 
subgroup of the semigroup (15). In other words, the approximation cascade (10)
is embedded into the approximation flow (15). The flow trajectory contains all
the cascade trajectory,
$$ \{ y_k(f)|\; k\in\Bbb{Z}_+\} \subset \{ y(t,f)|\; t\in\Bbb{R}_+\} . $$

Assume that the embedding of the cascade into the flow is smooth, in the 
sense that the derivative $\;dy/dt\;$ exists and is piecewise continuos on 
$\;\Bbb{R}_+\;$. Then the equation (13) can be rewritten in the differential 
form
\be
\frac{d}{dt}y(t,f) = v(y(t,f)) ,
\ee
in which the velocity
\be
v(f) \equiv \lim_{t\ra 0}\frac{\partial}{\partial t}y(t,f)
\ee
is a transformation $\;v:\;\Bbb{R}\ra\Bbb{T}(\Bbb{R})\;$ from $\;\Bbb{R}\;$ 
to a tangent space $\;\Bbb{T}(\Bbb{R})\;$. Integrating the evolution equation 
(16), we can cast it into the evolution integral
\be
\int_{y(t_1,f)}^{y(t_2,f)}\frac{dy}{v(y)} = t_2 - t_1 .
\ee
The fixed point of the approximation flow is defined as a zero of the velocity
\be
v(y^*(f)) = 0 .
\ee

The fourth step is the definition of {\it quasifixed points}, that is, of 
approximate fixed points. We cannot find fixed points exactly because the 
velocity of the approximation flow is not known, so we are not able to use 
(19). From another side, we cannot also define fixed points as the limit (12),
since an expression for $\;y_k(f)\;$ at arbitrary $\;k\ra\infty\;$ is not 
available.

To find approximate fixed points, or quasifixed points, we need to have an 
explicit form of the cascade velocity
\be
v_k(f) \equiv v(y(t,f)) \qquad (t\in [k,k+1]) .
\ee
We may define it as an Euler discretization of the derivative $\;dy/dt\;$. 
In doing this, we take into account that $\;F_k(g,u_k)\;$ depends on $\;k\;$ 
directly as well as through $\;u_k\;$. In this way, we may write the cascade 
velocity (20) as 
\be
v_k(f) \equiv V_k(g_k(f),u_k(g_k(f)))
\ee
with the finite difference
$$ V_k(g,u_k) = F_{k+1}(g,u_k) - F_k(g,u_k) + (u_{k+1}-u_k)
\frac{\partial}{\partial u_k}F_k(g,u_k) $$ 
and the coupling function $\;g_k\;$ given by the constraint (3). As the 
cascade trajectory approaches the fixed point, when $\;k\ra\infty\;$, then 
$\;v_k(f)\ra 0\;$. But to check whether the cascade velocity tends to zero, 
as $\;k\ra\infty\;$, we need to know the form of $\;F_k(g,u_k)\;$ for 
arbitrary $\;k\ra\infty\;$. If the latter would be known, then we could 
check directly the convergence condition
\be
F_{k+p}(g,u_{k+p}) - F_k(g,u_k) \ra 0 \qquad (k\ra \infty ) ,
\ee
in which $\;p\geq 1\;$. This condition would define those control functions 
which provide the convergence for the sequence $\;\{ f_k(g)\}_{k=0}^\infty\;$ 
of terms (1). Such a condition of defining control functions may be called 
the asymptotic fitting condition. It has been used for some anharmonic 
models [43-45]. However, the possibility of finding the general expression 
for $\;F_k(g,u_k)\;$ at the arbitrary $\;k\ra\infty\;$ is rather an extreme 
exception. Usually, just a few first terms are available only. Therefore, we 
need to have some working conditions for defining control functions in the 
case of finite numbers $\;k\;$ of $\;F_k\;$.

One such condition for defining control functions is the minimal--difference 
condition [18-25],
\be
F_{k+1}(g,u_k) - F_k(g,u_k) = 0 .
\ee
This condition makes zero only a part of the cascade velocity (21). Therefore 
(23) may be named the quasifixed--point condition.

Another quasifixed--point condition is the minimal--sensitivity condition 
[26-34],
\be
\frac{\partial}{\partial u_k}F_k(g,u_k) = 0 ,
\ee
which also makes zero only a part of (21).

A slight generalization of (24) following from (21) is the quasifixed--point 
condition [42]
\be
(u_{k+1}-u_k)\frac{\partial}{\partial u_k}F_k(g,u_k) = 0 .
\ee
The meaning of (25) is as follows: if (24) has a solution for $\;u_k\;$, then 
this solution gives the control function $\;u_k(g)\;$; when (24) has no 
solution for $\;u_k\;$, then we put $\;u_k=u_{k+1}\;$.

Defining the control functions from one of the quasifixed--point conditions, 
we obtain a sequence of quasifixed points $y_k(f)\;$ and, respectively, a 
sequence of their images $\;f_k(g)\;$ that are the sought approximations for 
$\;f(g)\;$. Several other types of quasifixed--point conditions have also 
been studied [49,50].

In the fifth step of the considered perturbation theory we find {\it corrected
approximations}. As far as the quasifixed--point conditions do not make the 
cascade velocity exactly zero, the trajectory does not stop at $\;y_k(f)\;$, 
though the motion slows down. If we accept the quasifixed--point condition 
(25), then the cascade velocity (21) in the vicinity of a quasifixed point 
$\;y_k(f)\;$ becomes
\be
v_k^*(f) = V_k^*(g_k(f),u_k(g_k(f))) , 
\ee
where
$$ V_k^*(g,u_k)\equiv F_{k+1}(g,u_k) - F_k(g,u_k) . $$
The motion in the interval of $\;t\in [k,k+1]\;$ is described by the 
approximation flow (15). Substituting into the evolution integral (18) the 
time limits $\;t_1 =k\;$ and $\;t_2=k+1\;$, we have
\be
\int_{y_k(f)}^{y_{k+1}(f)}\frac{dy}{v(y)} = 1 .
\ee
Using in (27), instead of $\;v(y)\;$, the velocity (26), we get the evolution 
integral
\be
\int_{y_k(f)}^{y_k^*(f)}\frac{dy}{v_k^*(y)} = 1
\ee
defining the corrected quasifixed point $\;y_k^*(f)\;$. Making in (28) the 
substitution $\;f\ra g_k^{-1}(g)\;$, we come, according to (7), to the 
integral
\be
\int_{f_k(g)}^{f_k^*(g)}\frac{df}{v_k^*(f)} = 1 ,
\ee
in which
\be
f_k^*(g) \equiv y_k^*(g_k^{-1}(g)) 
\ee
is the corrected $\;k\;$--order approximation.

\section{Stability of cascade}

A very important feature of the controlled perturbation theory is the 
possibility to control whether we are approaching the fixed point, that is, 
the correct answer, even if this exact answer is not known. This possibility 
is based on the semigroup property (9) of the approximation cascade, according
to which each point $\;y_k(f)\;$ is considered as a result of mapping from a 
previous point $\;y_p(f)\;$ with $\;p\leq k-1\;$. We are approaching an 
attracting fixed point if the mapping is contracting. This, of course, depends
on the perturbation algorithm and on the initial approximation. A given 
perturbation algorithm has a basin of attraction. The mapping can be 
contracting only if an initial approximation is in the basin of attraction. 
In iteration theory, the analog of the basin of attraction is the set of 
normality or Fatou set, while the complement of the Fatou set is called the 
Julia set. Attracting fixed points are in the Fatou set, while repelling fixed
points are in the Julia set. When the iteration is done by means of an entire 
transcendental function, then every point in the Julia set is a limit point of
repelling periodic points, that is, the Julia set is the closure of the set of
repelling periodic points [51].

Let us analyse when the mapping corresponding to the approximation cascade is 
contracting. Consider the change 
\be
y_p(f)\ra y_p(f)+\dt y_p(f)
\ee
with $\;\dt y_p(f)\ra 0\;$ and the variation
\be
\dt y_{kp}(f)\equiv y_k(y_p(f)+\dt y_p(f)) -y_k(y_p(f)) .
\ee
A particular case of (32) is
\be
\dt y_k(f)\equiv y_k(f+\dt f) -y_k(f) =\dt y_{k0}(f) ,
\ee
when the initial condition is changed.

To proceed further, we need to introduce the notation for mapping multipliers.
Define the {\it quasilocal multipliers}
\be
\mu_{kp}(f) \equiv \frac{\dt y_k(y_p(f))}{\dt y_p(f)}
\ee
and
\be
\mu_k(f)\equiv \frac{\dt y_k(f)}{\dt f} =\mu_{k0}(f) .
\ee
They satisfy the relation

\be
\mu_{kp}(f) =\mu_k(y_p(f)) 
\ee
and have the property
\be
\mu_{0k}(f)=\mu_0(f) = 1.
\ee
Other useful relations can be derived basing on the semigroup property
$$ y_{k+p}(f)=y_k(y_p(f))=y_p(y_k(f)) $$
and the variational derivative
$$ \frac{\dt y_k(f)}{\dt y_p(f)} = 
\frac{dy_k(f)/df}{dy_p(f)/df} = \frac{\mu_k(f)}{\mu_p(f)} . $$
In this way we obtain
\be
\mu_{kp}(f) =\frac{\mu_{k+p}(f)}{\mu_p(f)}
\ee
and
\be
\mu_{kp}(f)\mu_p(f) =\mu_{pk}(f)\mu_k(f) .
\ee

Introduce the {\it local multiplier}
\be
\mu^*_k(f)\equiv \frac{\dt y_k(f)}{\dt y_{k-1}(f)} = \mu_{1\;k-1}(f) .
\ee
Using (38), this can be also written as
\be
\mu_k^*(f) =\frac{\mu_k(f)}{\mu_{k-1}(f)} \qquad (k\geq 1) .
\ee
In the case of $\;k=1\;$,
$$ \mu_1^*(f) =\mu_1(f) . $$
The quasilocal multipliers (34) and (35) can be presented as products
$$ \mu_{kp}(f) =\prod_{j=p+1}^{k+p}\mu_j^*(f) , $$
\be
\mu_k(f) =\prod_{p=1}^{k}\mu_p^*(f)
\ee
of the local multipliers (40). In another form (42) reads
$$ \mu_{kp}(f) =\mu_{k+p}^*(f)\mu_{k-1\;p}(f) , $$
\be
\mu_k(f) =\mu_k^*(f)\mu_{k-1}(f) .
\ee

With these multipiers, the variation (32) becomes
\be
\dt y_{kp}(f) =\mu_{kp}(f)\dt y_p(f) .
\ee
Eq.(44) describes the deviation of the cascade trajectory at the 
$\;(k+p)\;$--step resulting from the variation $\;\dt y_p(f)\;$ at a 
$\;p\;$--step. The mapping, corresponding to (44), is contracting if the 
{\it condition of quasilocal contraction}
\be
|\mu_{kp}(f)| < 1
\ee
holds. This shows that the mapping is effectively contracting after $\;k\;$ 
steps starting from a $\;p\;$--step. Equivalently, one may say that the 
mapping is effectively contracting on the interval $\;[p,k+p]\;$.

In particular, for the interval $\;[0,k]\;$, we need to deal with the 
variation (33) which yields
\be
\dt y_k(f) =\mu_k(f)\dt f .
\ee
The condition of quasilocal contraction on the interval $\;[0,k]\;$ is
\be
|\mu_k(f)| < 1 ,
\ee
where $\;\mu_k(f)\;$ is the quasilocal multiplier (35) with $\;k\geq 1\;$.

If we are interested in the contraction property at just one step, from 
$\;k-1\;$ to $\;k\;$, then we have to consider the variation
\be
\dt y_k(f) =\mu_k^*(f)\dt y_{k-1}(f) ,
\ee
in which $\;\mu_k^*(f)\;$ is the local multiplier (40). The mapping is 
locally contracting at a $\;k\;$--step if the {\it condition of local 
contraction}
\be
|\mu_k^*(f)| < 1 
\ee
is valid.

The condition (49) is stronger than (47) in the following sense: if 
$\;|\mu_p^*|<1\;$ holds for all $\;p\in[0,k]\;$, then (47) follows from this 
because of the relation (42), though the inverse is not true. It may happen 
that (47) holds, but for some $\;p\;$ from the interval $\;[0,k]\;$ the 
condition (49) is not valid. In other words, there can exist the effective 
contraction on an interval $\;[0,k]\;$ although not for all steps there can 
be the local contraction. Simbolically, the relation between the two notions 
is:
$$ local\; contraction\ra quasilocal\; contraction . $$

The contraction for a mapping is the same as the stability for a cascade. The 
stability is characterized by Lyapunov exponents. Again, different kinds of 
such exponents can be defined. The {\it quasilocal Lyapunov exponent}
\be
\lm_{kp}(f) \equiv \frac{1}{k}\ln |\mu_{kp}(f)|
\ee
is related to the quasilocal multiplier (34), and the quasilocal exponent
\be
\lm_k(f) \equiv \frac{1}{k} \ln |\mu_k(f)| =\lm_{k0}(f) ,
\ee
to the multiplier (35). Taking into account (38), we have
\be
\lm_{kp}(f) =\frac{k+p}{k}\lm_{k+p}(f) - \frac{p}{k}\lm_p(f) .
\ee
The {\it local Lyapunov exponent}
\be
\lm_k^*(f) \equiv \ln |\mu_k^*(f)| =\lm_{1\;k-1}(f) 
\ee
is defined through the local multiplier (40). Because of (42), the quasilocal 
exponent (51) is an arithmetic average
\be
\lm_k(f) =\frac{1}{k}\sum_{p=1}^{k}\lm_p^*(f)
\ee
of the local exponents from (53).

The effective stability on the interval $\;[p,p+k]\;$ with $\;p\geq 0\;$ and 
$\;k\geq 1\;$, means that the {\it condition of quasilocal stability}
\be
\lm_{kp}(f) < 0
\ee
holds. In the case of $\;p=0\;$, this reduces to 
\be
\lm_k(f) < 0 .
\ee
As is evident, (55) and (56) follow from the contraction conditions (45) and 
(47), respectively.

The {\it condition of local stability} at a $\;k\;$--step, from $\;k-1\;$ to 
$\;k\;$, reads
\be
\lm_k^*(f) < 0 ,
\ee
resulting from (49). Anew, the condition (57) is stronger than (56) in the 
sense that the stability can exist on an interval but not necessarily at all 
points of the latter, while if (57) holds for all points of an interval, then 
(56) follows for this interval.

The maximal Lyapunov exponent, usually employed in dynamical theory, is
\be
\lm (f) \equiv \lim_{k\ra\infty}\lm_k(f) .
\ee
The {\it condition of asymptotic stability} implies that
\be
\lm (f) < 0 .
\ee
If the condition (56) is valid for all $\;k\geq 1\;$, then it is stronger than
(59). Thus, the relation between the different types of stability is as 
follows:
$$ local\; stability\ra quasilocal\; stability\ra asymptotic\; stability .$$

Recollect that the approximation--cascade trajectory 
$\;\{ y_k(f)\}_{k=0}^\infty\;$ is, by construction, bijective to the 
approximation sequence $\;\{ f_k(g)\}_{k=0}^\infty\;$. Each point $\;y_k(f)\;$
has its image $\;f_k(g)\;$ given by the relations (6) and (7). For the mapping
multipliers and Lyapunov exponents introduced above, we may  also define their
images as functions of $\;g\;$.

The image of the quasilocal multiplier (35) is
\be
M_k(g) \equiv \mu_k(g^{-1}_k(g)) .
\ee
This can also be written as
$$ M_k(g) =\mu_k(F_0(g,u_k(g)) =
\frac{\dt F_k(g,u_k(g))}{\dt F_0(g,u_k(g))} . $$
The image of the local multiplier (40) is
\be
M_k^*(g) \equiv \mu_k^*(g^{-1}_{k-1}(g)) .
\ee
Using the properties of multipliers, we may write
$$ M_k^*(g) =\mu_1(f_{k-1}(g)) =
\frac{\mu_k(F_0(g,u_{k-1}(g))}{\mu_{k-1}(F_0(g,u_{k-1}(g))} . $$
The contraction conditions (47) and (49) can be reformulated for the 
multipliers (60) and (61) giving
\be
|M_k(g)| < 1, \qquad |M_k^*(g)| < 1 ,
\ee
respectively.

For the image of the quasilocal Lyapunov exponent (51), we get
\be
\Lm_k(g) \equiv\lm_k(g^{-1}_k(g)) =\frac{1}{k}\ln |M_k(g)| ,
\ee
and for the image of the local Lyapunov exponent (53),
\be
\Lm_k^*(g) \equiv \lm_k^*(g^{-1}_{k-1}(g)) =\ln |M_k^*(g)| .
\ee
These are connected with each other, equivalently to  (54), through the 
arithmetic averaging
$$ \Lm_k(g) =\frac{1}{k}\sum_{p=1}^{k}\Lm_k^*(g) . $$
The stability conditions (56) and (57) can be written for the Lyapunov 
exponents (63) and (64), so that
\be
\Lm_k(g) < 0 , \qquad \Lm_k^*(g) < 0 , 
\ee
respectively.

As far as the stability conditions for the Lyapunov exponents are a 
reformulation of the contraction conditions for the mapping multipliers, 
in what follows we shall often refer to any of them as to the stability 
conditions.

\section{Stability of flow}

The stability of motion analized above concerns the stability of an 
approximation cascade. If the latter is embedded into an approximation flow, 
we need to check the stability of the flow as well. This can be done by making
in the evolution equation (16) 

the substitution
\be
y(t,f)\ra y(t,f)+\dt y(t,f)
\ee
implying that $\;\dt y(t,f)\ra 0\;$. Then we find
\be
\dt y(t,f) =\dt y(t_0,f)\exp\left \{ \int_{t_0}^{t}\lm(t',f)dt'\right \} ,
\ee
where $\;t_0\leq t\;$, and
\be
\lm(t,f) \equiv \frac{\dt v(y(t,f))}{\dt y(t,f)}
\ee
is the local Lyapunov exponent for the flow. According to (20), the flow 
velocity in the time interval $\;k\leq t\leq k+1\;$ is given by the 
corresponding cascade velocity. For this reason, we put $\;t_0 =k\;$ in (67) 
and get
\be
\dt y(t,f) =\dt y_k(f)\exp \{ \stackrel{-}{\lm}_k(f)(t-k)\} ,
\ee
where we took into account that, in compliance with (14), $\;y(k,f)=y_k(f)\;$,
and that (68) gives
\be
\stackrel{-}{\lm}_k(f) \equiv \frac{\dt v_k(y_k(f))}{\dt y_k(f)} .
\ee
The approximation flow near a quasifixed point $\;y_k(f)\;$ is stable if
\be
\stackrel{-}{\lm}_k(f) < 0 .
\ee
The image of (70) is
\be
\stackrel{-}{\Lm}_k(g) \equiv 
\stackrel{-}{\lm}_k(g_k^{-1}(g)) =\frac{\dt v_k(f_k(g))}{\dt f_k(g)} ,
\ee
and the condition (71) of the local stability of the flow acquires the form
\be
\stackrel{-}{\Lm}_k(g) < 0 .
\ee

Note that the local Lyapunov exponents for a cascade and for a flow, into 
which the cascade is embedded, are, generally, different. This means that 
(53) does not coincide with (70). Therefore, it may happen that the cascade 
at a point $\;y_k(f)\;$ is locally stable but the flow at the same point is 
not, or vice versa. To understand better the distinction between (70) and 
(53), we may invoke the definition of the cascade velocity as of the finite 
difference 
\be
v_k(f) = y_{k+1}(f) - y_k(f) .
\ee
Then (70) becomes
\be
\stackrel{-}{\lm}_k(f) =\mu_{k+1\; k}(f) - \mu_{kk}(f) .
\ee
Using the properties of the quasilocal multipliers (34), we may transform 
(75) to
$$ \stackrel{-}{\lm}_k(f) =\mu_{kk}(f)\left [ \mu^*_{2\;k+1}(f) -1
\right ] . $$
According to (50),(51) and (53), we have   
$$ |\mu_{kk}(f)| =\exp \left \{ \sum_{p=k+1}^{2k}\lm_p^*(f)\right \} , $$
$$ |\mu_k(f)| =\exp\{ k\lm_k(f) \} , $$
$$ |\mu_k^*(f)| =\exp \{ \lm_k^*(f)\} . $$
As we see, there is no simple relation between (75) and (53).

\section{Stability and convergence}

Since the cascade trajectory is bijective to the approximation sequence, 
the stability conditions for the cascade should characterize the corresponding
convergence properties for the sequence.

The deviation of the trajectory point $\;y_k(f)\;$ from the fixed point 
$\;y^*(f)\;$ is
\be
\Delta y_k(f) \equiv y_k(f) - y^*(f) .
\ee
Consider the deviation $\;\Delta y_{k+p}(f)\;$ assuming that $\;y_p(f)\;$ is 
close to $\;y^*(f)\;$ in the sense that
\be
|y_p(f) - y^*(f) | \ll | y^*(f)| .
\ee
Employing the definition of the fixed point,
$$ y_k(y^*(f)) = y^*(f) , $$
we find
\be
\Delta y_{k+p} (f)\simeq \mu_{kp}(f)\Delta y_p(f) .
\ee
From (78), it is easy to derive the equations
$$ \Delta y_k(f)\simeq \mu_k(f)\Delta f , $$
\be
\Delta y_k(f) \simeq \mu_k^*(f)\Delta y_{k-1}(f) ,
\ee
in which $\;\Delta f \equiv f - y^*(f)\;$. In its turn, (79) gives
$$ |\Delta y_k(f)| \simeq |\Delta f|\exp\{k\lm_k(f)\} , $$
\be
|\Delta y_k(f)| \simeq |\Delta y_{k-1}(f)|\exp \{\lm_k^*(f)\} .
\ee

The image of (76) is
\be
\Delta f_k(g) \equiv f_k(g) - f(g) .
\ee
Respectively, (80) yields
$$ |\Delta f_k(g)| \simeq |\Delta f_0(g)|\exp \{ k\Lm_k(g)\} . $$
\be
|\Delta f_k(g)| \simeq |\Delta f_{k-1}(g)|\exp \{\Lm_k^*(g)\} .
\ee
The accuracy of an approximation $\;f_k(g)\;$, as compared to the exact value
$\;f(g)\;$, is characterized by the absolute error (81). When the condition of
quasilocal stability (56) holds, that is when $\;\Lm_k(g)<0\;$, then (82) 
shows that the accuracy of $\;f_k(g)\;$ is better than that of $\;f_0(g)\;$, 
although the accuracy may decrease with respect to $\;f_{k-1}(g)\;$. If 
$\;\Lm_k(g) <0\;$ for all $\;k\geq 1\;$, then the error (81) tends to zero, 
which means the convergence of the sequence $\;\{f_k(g)\}_{k=0}^\infty\;$. 
When the conditions of local stability (57) is valid, then $\;\Lm_k^*(g)<0\;$,
and (82) shows that the accuracy of $\;f_k(g)\;$ is better than that of 
$\;f_{k-1}(g)\;$. If the condition of local stability occurs for all 
$\;k\geq 1\;$, then the accuracy of approximations improves at each step. 
This also means the convergence of the sequence 
$\;\{ f_k(g)\}_{k=0}^\infty\;$, since (57) is stronger than (56).

Thus, the stability conditions (65) are  sufficient conditions for the 
convergence of $\;\{ f_k(g)\}_{k=0}^\infty\;$. The necessary and sufficient 
condition of convergence is
\be
k\Lm_k(g)\ra -\infty \qquad (k\ra \infty ) .
\ee
For the corresponding quasilocal exponent (51), this reads
\be
k\lm_k(f)\ra -\infty \qquad (k\ra \infty ) ,
\ee
which is equivalent to the asymptotic condition
\be
|\mu_k(f)|\ra 0 \qquad (k\ra\infty) .
\ee
The convergence conditions (83)--(85) are weaker than the conditions of 
quasilocal stability (56), local stability (57) and asymptotic stability (59).
The maximal Lyapunov exponent (58) can be zero; nevertheless, the convergence 
of the approximation sequence will persist provided (84) is valid.

By analogy with the usage of the terms quasilocal or local contraction, as 
applied to a mapping, and quasilocal or local stability, as applied to a 
cascade, we may use the terms quasilocal or local convergence, as applied 
to a sequence. We shall say that a sequence of approximations is {\it 
quasilocally convergent} on the interval $\;[0,k]\;$ if the condition of 
quasilocal stability (56), i.e., $\;\Lm_k(g)<0\;$, holds on this interval. 
The sequence will be named {\it locally convergent} at a $\;k\;$--step if the 
condition of local stability (57), that is, $\;\Lm_k^*(g)<0\;$, is valid at 
this given $\;k\;$. When the conditions of quasilocal and local convergence 
hold everywhere, that is, are true for all $\;k\geq 1\;$, then they are 
stronger than the convergence criterion (63). There is the following relation 
between different notions of convergence:
$$ local\; convergence\ra quasilocal\; convergence\ra convergence .$$ 
This notions should not be confused with the point convergence, convergence 
on an interval and uniform convergence of the sequence 
$\;\{ f_k(g)\}_{k=0}^\infty\;$ with respect to the variable $\;g\;$.

Because of the exponential renormalization of the deviation in (82), one may 
say, when the corresponding convergence conditions are valid, that there 
occurs an {\it exponential convergence}.

In order to find out when the corrected approximation $\;f_k^*(g)\;$ is better
that $\;f_k(g)\;$, consider the deviation
\be
\Delta y_k^*(f) \equiv y_k^*(f) -y^*(f) .
\ee  
From the evolution equation (16) we obtain
\be
\Delta y_k^*(f)\simeq \Delta y_k(f)\exp \{ \stackrel{-}{\lm}_k(f)\} ,
\ee
where $\; \stackrel{-}{\lm}_k(f)\;$ is defined in (70). The image of (86) is
\be
\Delta f^*_k(g) \equiv f_k^*(g) - f(g) .
\ee
Whence, (87) gives
\be
|\Delta f_k^*(g)|\simeq |\Delta f_k(g)|\exp \{ \stackrel{-}{\Lm}_k(g)\} .
\ee
Eq.(89) shows that the corrected approximation $\;f_k^*(g)\;$ is more accurate
than $\;f_k(g)\;$ when the approximation flow is locally stable, so that 
condition (73) holds.

Let us also observe when the corrected approximation $\;f_k^*(g)\;$ is better 
than $\;f_{k+1}(g)\;$. This happens when
$$ |\Delta y_k^*(f)| < |\Delta y_{k+1}(f)| . $$
The latter inequality leads to
$$ \stackrel{-}{\lm}_k(f) <\lm_{k+1}^*(f) , $$
\be
\stackrel{-}{\Lm}_k(g) < \Lm_{k+1}^*(g) .
\ee
When the approximation cascade is locally stable at the $\;(k+1)\;$--step, 
then the improvement of the accuracy for $\;f_k^*(g)\;$, as compared to 
$\;f_k(g)\;$, can be achieved only if the enveloping approximation flow is 
stable at its $\;k\;$--point with the local exponent satisfying (90). If the 
approximation cascade is locally unstable, so that $\;\lm_{k+1}^*(f)>0\;$, 
then the corrected approximation $\;f_k^*(g)\;$ can be of much better accuracy
than $\;f_k(g)\;$ even in the case of an unstable approximation flow, provided
(90) is valid. In the latter case, to easier satisfy (90), the motion should 
be damped, so that to make $\;\stackrel{-}{\lm}_k(f)\;$ smaller. This can be 
done by incorporating into the  definition of the cascade velocity (21) a 
damping parameter $\;\dt_k\;$ lowering $\;v_k(f)\;$,
$$ v_k(f) \ra \dt_kv_k(f) . $$
The value of the damping parameter $\;\dt_k\;$ can be found from additional 
conditions. For example, one may require the coincidence of some asymptotic 
values for the corrected approximation $\;f_k^*(g)\;$ and for the exact 
$\;f(g)\;$, of course, if such asymptotic values of $\;f(g)\;$ are available 
[39,49]. Another option is to put $\;\dt_k=\frac{1}{2}\;$, which corresponds 
to diminishing twice the step of the calculational procedure.

The possibility of improving the accuracy even for unstable approximation 
cascades and flows is the main advantage of the corrected approximations.

\section{Anharmonic oscillator}

To illustrate the ideas of the approach we choose an anharmonic--oscillator 
model. The anharmonicity of oscillations plays a very important role in many 
physical problems, for instance, in anharmonic crystals [52,53].

Suppose we need to find the energy levels of an anharmonic oscillator with the
Hamiltonian
\be
H =-\frac{1}{2}\frac{d^2}{dx^2} +\frac{1}{2}x^2 + gx^4 ,
\ee
in which $\;x\in(-\infty,\infty)\;$ and the coupling, or anharmonicity, 
parameter $\;g\geq 0\;$. Note that several problems of quantum mechanics can 
be reduced to oscillator--type models by a special change of variables 
[54,55].

Emphasize that our aim here is not simply the calculation of the energy levels
but the demonstration how the controlled perturbation theory, formulated as 
dynamical theory, works for such a touchstone model as (91). In our previous 
papers [38-40] we have considered solely the first step of the theory as 
applied to an anharmonic oscillator. One step, of course, does  not permit 
yet to illustrate and exploit in full the underlying ideas. This is why we 
have again to turn our attention to the model (91) extending the consideration
to the higher--order approximations.

It is natural to start from the harmonic oscillator whose Hamiltonian
\be
H_0 = -\frac{1}{2}\frac{d^2}{dx^2} +\frac{1}{2}u^2x^2
\ee
contains an unknown parameter $\;u\;$. For convenience, we introduce the 
notation
\be
E_k(g,u) \equiv \left ( n+\frac{1}{2}\right )F_k(g,u)
\ee
for the $\;k\;$--order approximation of the spectrum. The quantum index 
$\;n=0,1,2,\ldots \;$ in $\;E_k\;$ and $\;F_k\;$ is not written explicitly 
for the sake of brevity.

The sequence $\;\{ F_k(g,u)\}_{k=0}^\infty\}\;$ is to be obtained by the 
Rayleigh--Schr\"odinger perturbation theory starting from
\be
F_0(g,u) = u .
\ee
In what follows we shall need the notation 
$$ \al =\al(u) \equiv 1 -\frac{1}{u^2} , $$
\be
\bt =\bt(u) \equiv \frac{6\ga g}{u^3} , 
\ee
$$\ga\equiv \frac{n^2+n+1/2}{n+1/2} . $$
The first four approximations, under a fixed $\;u\;$, are
$$ F_1(g,u) = u-\frac{u}{4}\left ( 2\al -\bt\right ) , $$
$$ F_2(g,u) = F_1(g,u) -\frac{u}{8}\left ( \al^2 -2\al\bt +2a\bt^2
\right ) , $$
$$ F_3(g,u) = F_2(g,u) -\frac{u}{16}\left ( \al^3 -4\al^2\bt +10a
\al\bt^2 -3b\bt^3\right ) , $$
\be
F_4(g,u) = F_3(g,u) -\frac{u}{32}\left (\frac{5}{4}\al^4 -8\al^3\bt +
35a\al^2\bt^2-24b\al\bt^3+4c\bt^4 \right ) ,
\ee
in which
$$ a=a(\ga)\equiv \frac{17n^2+17n+21}{(6\ga)^2} , $$
$$ b=b(\ga) \equiv 
\frac{125n^4+250n^3+472n^2+347n+111}{(n+1/2)(6\ga)^3} , $$
$$ c=c(\ga)\equiv 
\frac{10689n^4+21378n^3+60616n^2+49927n+30885}{8(6\ga)^4}. $$

From the quasifixed--point condition (25) of the form
$$ \frac{\partial}{\partial u_1}F_1(g,u_1) =0 $$
we get the equation
\be
u_1^3 - u_1 -6\ga g = 0
\ee
for the control function $\;u_1(g)\;$. Eqs.(95) and (97) tell us that
$$ \al(u_1)=\bt(u_1) . $$
Introducing the notation
\be 
\al_k\equiv \al(u_k) = 1 -\frac{1}{u_k^2} , \qquad \bt_k\equiv \bt(u_k) ,
\ee
we have
\be
F_{k+1}(g,u_1) = F_k(g,u_1)+u_1A_{1k}\al_1^{k+1} \qquad ( k=0,1) ,
\ee
where
$$ A_{10} =-\frac{1}{4}, \qquad A_{11} =\frac{1}{8}\left ( 1 -2a
\right ) . $$

The equation $\;\partial F_2(g,u_2)/\partial u_2 =0\;$ has no real solutions 
for $\;u_2\;$, therefore, according to (25), we put
\be
u_2(g) = u_3(g) ,
\ee
and $\;u_3\;$ being defined by the equation
$$ \frac{\partial}{\partial u_3}F_3(g,u_3) = 0 . $$
The latter yields
\be
u_3^3 - u_3 -6\ga_3 g = 0 
\ee
with
\be
\ga_3\equiv \kp\ga , \qquad \al_3=\kp\bt_3
\ee
and with $\;\kp\;$ given by the equation
\be
5\kp^3-24\kp^2+70a\kp-24b=0 .
\ee

Eqs.(101)--(103) make it possible to find
\be
F_{k+1}(g,u_3) = F_k(g,u_3) + u_3A_{3k}\al_3^{k+1} \qquad (k\leq 3) ,
\ee
where
$$ A_{30} =-\frac{1}{4}\left ( 2-\frac{1}{\kp}\right ) , $$
$$ A_{31} = -\frac{1}{8}\left ( 1 -\frac{2}{\kp} + 2a_3\right ) , $$
$$ A_{32} = -\frac{1}{16} \left ( 1 -\frac{4}{\kp} +10a_3 - 3b_3\right ) , $$
$$ A_{33} = -\frac{1}{32} \left ( \frac{5}{4} -\frac{8}{\kp} + 35a_3 -24b_3 +
4c_3\right ) , $$
and
$$ a_3 \equiv a(\ga_3) , \qquad b_3\equiv b(\ga_3) , \qquad 
c_3\equiv c(\ga_3). $$

Definition (3) of the coupling function $\;g_k(f)\;$ because of (94), gives 
the equation
\be
u_k(g_k(f)) = f .
\ee
From here we find
\be
g_k(f) =\frac{f(f^2-1)}{6\ga_k} ,
\ee
where we took into account that
\be
\al(u_k(g_k(f))) = 1 -\frac{1}{f^2}
\ee
and used the notation
$$ \ga_1\equiv\ga , \qquad \ga_2\equiv\ga_3 =\kp\ga . $$

For the cascade velocity (26) we get
\be
v_k^*(f) = A_kf\left ( 1 -\frac{1}{f^2}\right )^{k+1}
\ee
with
$$ A_k\equiv A_{kk} , \qquad A_{2k} \equiv A_{3k} . $$
The evolution integral (29) becomes
\be
\int_{f_k(g)}^{f_k^*(g)}\frac{f^{2k+1}df}{(f^2-1)^{k+1}} = A_k .
\ee
With the notation
\be
f_k^*(g)\equiv \sqrt{1+z_k^*(g)} , \qquad f_k(g) \equiv \sqrt{1+z_k(g)} ,
\ee integral (109) transforms to 
\be
\int_{z_k(g)}^{z_k^*(g)}\frac{(1+z)^k}{z^{k+1}}dz = 2A_k .
\ee
Employing the binomial formula
$$ (1+z)^k=\sum_{p=0}^{k}C_k^p z^p ; \qquad C_k^p\equiv 
\frac{k!}{(k-p)!p!} , $$
we integrate (111) obtaining the equation
\be
z_k^* = z_k\exp \left \{ \sum_{p=0}^{k-1}\frac{C_k^p}{k-p}\left [ 
\frac{1}{(z_k^*)^{k-p}} - \frac{1}{z_k^{k-p}}\right ] + 2A_k\right\} ,
\ee
in which
$$ z_k^* = z_k^*(g), \qquad z_k = z_k(g) . $$
Introducing the polynomial
$$ P_k(x) \equiv \sum_{p=0}^{k-1}\frac{C_k^p}{k-p} x^{k-p} $$
with
$$ P_1(x) = x, $$
$$ P_2(x) = 2x+\frac{1}{2}x^2 , $$
$$ P_3(x) = 3x+\frac{3}{2}x^2 +\frac{1}{3}x^3 , $$
we can cast (112) into a more compact form
\be
z_k^* = z_k\exp\left \{ P_k\left (\frac{1}{z_k^*}\right ) -
P_k\left ( \frac{1}{z_k}\right ) + 2A_k\right \} .
\ee

In this way, the $\;k\;$--order approximation for the spectrum of the 
Hamiltonian (91), i.e.
$$ e_k(g) \equiv E_k(g,u_k(g)) , $$
owing to notation (93) can be written as
\be
e_k(g) =\left ( n  +\frac{1}{2}\right ) f_k(g) = e_k(n,g)
\ee
with $\;f_k(g) \equiv F_k(g,u_k(g))\;$. The corrected $\;k\;$--order 
approximation is
\be
e_k^*(g) =\left ( n+\frac{1}{2}\right ) f_k^*(g) = e_k^*(n,g) .
\ee
The accuracy of these approximations is characterized by their percentage 
errors
$$ \ep_k(g) \equiv \frac{e_k(g)-e(g)}{e(g)}\cdot 100\% , $$
$$ \ep_k^*(g) \equiv \frac{e_k^*(g)-e(g)}{e(g)}\cdot 100\% , $$
with respect to exact numerical values $\;e(g)\;$.

Table I illustrates the accuracy of the approximations
$$ e_1(g)=E_1(g,u_1(g)); \qquad e_2(g) =E_2(g,u_3(g)) ; $$
$$ e_3(g) =E_3(g,u_3(g)); \qquad e_4(g) =E_4(g,u_3(g)) , $$
and table II describes the accuracy of the corresponding approximations
$$ e_1^*(g) =E_1^*(g,u_1(g)); \quad e_2^*(g) =E_2^*(g,u_3(g)); 
\quad e_3^*(g)=E_3^*(g,u_3(g)) , $$
where $\;E_k^*\;$ means the right--hand side of (115). The errors 
$\;\ep_k(g) \equiv\ep_k(n,g)\;$ and $\;\ep_k^*(g)\equiv\ep_k^*(n,g)\;$ depend 
on the value of the coupling parameter $\;g\in\Bbb{R}_+\;$ and on the level 
number $\;n\in\Bbb{Z}_+\;$. The uniform accuracy of an approximation may be 
characterized by the maximal error
$$ \ep_k\equiv \sup_{g\in\Bbb{R}_+}\sup_{n\in\Bbb{Z}_+}|\ep_k(n,g)| $$
or, respectively by
$$  \ep_k^*\equiv \sup_{g\in\Bbb{R}_+}\sup_{n\in\Bbb{Z}_+}|\ep_k^*(n,g)| . $$
From the table I we have
$$ \ep_1=2.0\% , \quad \ep_2=0.45\%, \quad \ep_3=0.84\%, \quad 
\ep_4=0.50\% , $$
and from the table II,
$$ \ep_1^*=0.40\%, \quad \ep_2^*=0.37\% , \quad \ep_3^*=0.65\% . $$
As is seen, $\;\ep_k^* <\ep_k\;$, which means that the corrected approximation
(115) improves the accuracy of (114). However, the error do not monotonically 
decrease as the approximation order $\;k\;$ increases. This should be related 
to the occurrence of local instabilities in the calculational procedure, which
can be detected by the stability analysis.

To analyse the stability, we, first, need to define the trajectory of the 
approximation cascade, whose points are given by (6). In the considered case 
we find
\be
y_k(f) =f+f\sum_{p=1}^{k}B_{kp}\left ( 1 -\frac{1}{f}\right )^p
\ee
with the coefficients
$$ B_{11} = A_{10} , $$
$$ B_{21}=A_{10}, \quad B_{22}=A_{11} , $$
$$ B_{31}=A_{30}, \quad B_{32}=A_{31} , \quad B_{33}=A_{32} , $$
$$ B_{41}=A_{30},\quad B_{42}=A_{31},\quad B_{43}=A_{32},\quad 
B_{44}=A_{33} . $$
For the quasilocal multipliers (35) we obtain
\be
\mu_k(f) = 1+\sum_{p=1}^{k}B_{kp}\left ( 1 +\frac{p-1}{f}\right )
\left ( 1 -\frac{1}{f}\right )^{p-1} .
\ee
The local multipliers (40) can be found from (117) by means of (41).

The numerical analysis shows that the condition of quasilocal contraction (47)
holds true,
\be
|\mu_k(f)|< 1 \qquad (k=1,2,3,4),
\ee
and, respectively, the condition of quasilocal stability (56) is valid. This 
means that all approximations (114), with $\;k\geq 1\;$ are closer to the 
exact values than the zero approximation. But the condition of local 
contraction (49) does not necessary holds for all $\;f\in(1,\infty )\;$ and 
$\;n\in\Bbb{Z}_+\;$. Because of this the accuracy of approximations may 
not improve at each step.

The local Lyapunov exponent (70) for the approximation flow is
\be
\stackrel{-}{\lm}_k(f) = A_k\left ( 1 +\frac{2k+1}{f^2}\right )
\left ( 1 -\frac{1}{f^2}\right )^k .
\ee
Since $\;f\in (1,\infty)\;$, the sign of (119) is defined by that of 
$\;A_k\;$,
$$ {\rm sgn}\stackrel{-}{\lm}_k(f) ={\rm sgn} A_k . $$
For $\;A_k\;$ we have the inequalities
$$ -\frac{1}{48}\leq A_1\leq \frac{1}{144} , $$
$$ -0.002496 \leq A_2 \leq 0.003001 , $$
$$ -0.000525 \leq A_3 \leq -0.000439 $$
depending on the quantum number $\;n=0,1,2,\ldots \;$. Therefore (119) yields
\be
\stackrel{-}{\lm}_1(f)< 0.007, \quad \stackrel{-}{\lm}_2(f)< 0.003,
\quad \stackrel{-}{\lm}_3(f)< 0 . 
\ee
Thus, at the first two steps the local flow exponents (119) become positive 
for some energy levels. The positiveness of the local Lyapunov exponents 
signifies the occurrence of local chaos. However, at the third step the motion
stabilizes, since $\;\stackrel{-}{\lm}_3(f)<0\;$ for all $\;f\in(1,\infty)\;$ 
and $\;n\in\Bbb{Z}_+\;$. The same inequalities (120) are valid for the images 
of $\;\stackrel{-}{\lm}_k(f)\;$ given by (72), that is, for 
$\;\stackrel{-}{\Lm}_k(g)\;$ as functions of $\;g\in\Bbb{R}_+\;$ and 
$\;n\in\Bbb{Z}_+\;$.

\section{Conclusion}

Perturbation theory can be made convergent by introducing control functions. 
The reformulation of perturbation theory to the language of dynamical theory 
makes it possible to control convergence of the approximation sequence by 
checking stability conditions. The sequence of perturbative approximations is 
bijective to the trajectory of the approximation cascade. The control 
functions are defined from quasifixed--point conditions. The terms of the 
perturbation sequence are images of quasifixed points. Each quasifixed point 
can be corrected by embedding the approximation cascade into an approximation 
flow and considering the motion near the given quasifixed point.

The stability of calculational procedure is controlled by mapping multipliers 
and Lyapunov exponents. Several such characteristics can be introduced each of
them being responsible for controlling particular stability properties. The 
quasilocal multipliers control the quasilocal contraction of a mapping on 
intervals. The local multipliers control the local contraction of a mapping 
at each step. The quasilocal Lyapunov exponents control the quasilocal 
stability of the approximation cascade on intervals. The local Lyapunov 
exponents control the local stability of the approximation cascade at each 
point of its trajectory. The classical Lyapunov exponents describe the 
asymptotic stability of the approximation cascade. The local Lyapunov 
exponents for the approximation flow, enveloping the approximation cascade, 
define the local stability of motion near the corresponding quasifixed points.
The stability conditions are directly related to the character of convergence 
of the approximation sequence. Because of the possibility to control the 
convergence by means of control functions, mapping multipliers and Lyapunov 
exponents, the developed approach is called the controlled perturbation 
theory.

As an illustration of the method, the calculation of energy levels for a 
one--dimensional anharmonic oscillator is accomplished. Higher--order 
approximations are considered, as compared to our previous papers. The 
stability analysis showed that, in this case, the approximation cascade is 
quasilocally stable but not locally stable. In other words, the approximation 
sequence is quasilocally convergent but not locally convergent. This results 
in local chaos at the beginning of the cascade trajectory and, respectively, 
in a nonmonotonic fluctuation of errors for several first approximations. The 
local chaos can, in principle, be suppressed by introducing a damping 
parameter diminishing the cascade velocity. However, here we did not study 
the mechanism of such a suppression of chaos. We hope to do this in future 
papers.

\newpage

{\bf TABLE I}

\vspace{0.5cm}

The accuracy of the approximations $\;e_k(n,g)\;$ of the controlled 
perturbation theory for the energy levels of the one--dimensional anharmonic 
oscillator, as compared to the exact numerical values $\;e(n,g)\;$.

\vspace{1cm}

\begin{center}

\begin{tabular}{|c|c|c|c|c|c|c|}\hline
$g$&$n$&$e(n,g)$&$\ep_1(n,g)$&$\ep_2(n,g)$&$\ep_3(n,g)$&$\ep_4(n,g)$ \\ \hline
      &  0 & 0.50726 & 0.006  & $ 10^{-4}$& $ 10^{-4}$&  $ 10^{-4}$ \\
      &  1 & 1.5357  & 0.009  &  0.001    & $-10^{-4}$&  $-10^{-4}$ \\
      &  2 & 2.5909  & 0.001  &  0.008    &  --0.001  &  $-10^{-4}$   \\
      &  3 & 3.6711  & --0.008 &  0.017   &  --0.003  &   --0.001 \\
0.01  &  4 & 4.7749  & --0.019 &  0.027   &  --0.006  &   --0.002 \\
      &  5 & 5.9010  & --0.030 & --0.002  & $ 10^{-4}$&  $-10^{-4}$ \\
      &  6 & 7.0483  & --0.040 & --0.001  &   0.001   &  $ 10^{-4}$ \\
      &  7 & 8.2158  & --0.051 & --0.001  &   0.001   &  $ 10^{-4}$ \\
      &  8 & 9.4030  & --0.066 & --0.004  &  --0.003  &  --0.003   \\ \hline
      &  0 & 0.63799 &  0.57  &  0.10     &   0.01    &  --0.003  \\
      &  1 & 2.0946  &  0.49  &  0.15     &  --0.01   &  --0.004   \\
      &  2 & 3.8448  & --0.06  &  0.21    &  --0.24   &  --0.13   \\
      &  3 & 5.7970  & --0.30  &  0.20    &  --0.36   &  --0.20   \\
0.3   &  4 & 7.9100  & --0.40  &  0.23    &  --0.40   &  --0.21   \\
      &  5 & 10.167  & --0.51  & --0.09   &   0.01    &  --0.01   \\
      &  6 & 12.540  & --0.53  & --0.02   &   0.05    &    0.03    \\
      &  7 & 15.030  & --0.59  & --0.01   &   0.03    &    0.01   \\
      &  8 & 17.620  & --0.63  & --0.01   &   0.03    &    0.01   \\ \hline
      &  0 & 0.80377 &  1.09  &  0.21     &   0.02    &  --0.01   \\
      &  1 & 2.7379  &  0.81  &  0.25     &  --0.03   &  --0.01   \\
      &  2 & 5.1780  & --0.11  &  0.25    &  --0.40   &  --0.22   \\
      &  3 & 7.9400  & --0.40  &  0.25    &  --0.51   &  --0.28   \\
 1    &  4 & 10.960  & --0.55  &  0.24    &  --0.58   &  --0.32   \\
      &  5 & 14.203  & --0.66  & --0.12   &   0.03    &  --0.003   \\
      &  6 & 17.630  & --0.69  & --0.05   &   0.05    &    0.01   \\
      &  7 & 21.240  & --0.77  & --0.06   &   0.01    &  --0.03   \\
      &  8 & 25.000  & --0.80  & --0.05   &  --0.002  &  --0.03   \\ \hline
      &  0 & 3.9309  &  1.97  &  0.44     &   0.04    &  --0.03   \\
      &  1 & 14.059  &  1.24  &  0.39     &  --0.05   &  --0.02   \\
      &  2 & 27.550  & --0.24  &  0.22    &  --0.67   &  --0.41   \\
      &  3 & 43.010  & --0.61  &  0.18    &  --0.79   &  --0.48   \\
200   &  4 & 60.030  & --0.75  &  0.19    &  --0.82   &  --0.48   \\
      &  5 & 78.400  & --0.85  & --0.18   &   0.03    &  --0.02   \\
      &  6 & 97.900  & --0.88  & --0.10   &   0.03    &  --0.02   \\
      &  7 & 118.40  & --0.88  & --0.03   &   0.05    &    0.01   \\
      &  8 & 139.90  & --0.92  & --0.03   &   0.03    &  --0.01   \\ \hline
      &  0 & 18.137  &  2.01  &  0.45     &   0.04    &  --0.03   \\
      &  1 & 64.987  &  1.26  &  0.40     &  --0.05   &  --0.02   \\
      &  2 & 127.51  & --0.25  &  0.21    &  --0.69   &  --0.43   \\
      &  3 & 199.20  & --0.64  &  0.16    &  --0.82   &  --0.50   \\
20000 &  4 & 278.10  & --0.76  &  0.18    &  --0.84   &  --0.49   \\
      &  5 & 363.20  & --0.84  & --0.16   &   0.05    &  --0.01   \\
      &  6 & 454.00  & --0.95  & --0.16   &  --0.04   &  --0.09   \\
      &  7 & 548.90  & --0.91  & --0.05   &   0.03    &  --0.01   \\
      &  8 & 648.50  & --0.93  & --0.03   &   0.03    &  --0.01   \\ \hline
\end{tabular}

\end{center}

\newpage

{\bf TABLE II}

\vspace{0.5cm}

The accuracy of the corrected approximations $\;e_k^*(n,g)\;$ for the energy 
levels of the one--dimensional anharmonic oscillator, compared to the exact 
values $\;e(n,g)\;$.

\vspace{1cm}

\begin{center}

\begin{tabular}{|c|c|c|c|c|c|c|}\hline
$g$&$n$&$e(n,g)$&$\ep_1^*(n,g)$&$\ep_2^*(n,g)$&$\ep_3^*(n,g)$ \\ \hline
      &  0 & 0.50726 & 0.005   & $ 10^{-4}$ & $ 10^{-4}$  \\
      &  1 & 1.5357  & 0.007   &   0.001    & $-10^{-4}$  \\
      &  2 & 2.5909  & 0.001   &   0.008    &  --0.001  \\
      &  3 & 3.6711  & --0.006 &   0.017    &  --0.003  \\
0.01  &  4 & 4.7749  & --0.013 &   0.027    &  --0.006  \\
      &  5 & 5.9010  & --0.021 & --0.002    & $ 10^{-4}$  \\
      &  6 & 7.0483  & --0.028 & --0.001    &    0.001     \\
      &  7 & 8.2158  & --0.035 & $-10^{-4}$ &    0.001     \\
      &  8 & 9.4030  & --0.046 & --0.004    &  --0.003     \\ \hline
      &  0 & 0.63799 &   0.25  &   0.08     &    0.01     \\
      &  1 & 2.0946  &   0.19  &   0.11     &  --0.01     \\
      &  2 & 3.8448  & --0.11  &   0.11     &  --0.22     \\
      &  3 & 5.7970  & --0.20  &   0.07     &  --0.33     \\
0.3   &  4 & 7.9100  & --0.21  &   0.06     &  --0.36     \\
      &  5 & 10.167  & --0.26  & --0.04     &    0.004     \\
      &  6 & 12.540  & --0.23  &   0.02     &    0.04     \\
      &  7 & 15.030  & --0.26  &   0.01     &    0.02     \\
      &  8 & 17.620  & --0.27  &   0.01     &    0.02     \\ \hline
      &  0 & 0.80377 &   0.28  &   0.16     &    0.01     \\
      &  1 & 2.7379  &   0.19  &   0.16     &  --0.02     \\
      &  2 & 5.1780  & --0.19  &   0.03     &  --0.35     \\
      &  3 & 7.9400  & --0.24  & --0.04     &  --0.44     \\
 1    &  4 & 10.960  & --0.26  & --0.08     &  --0.49     \\
      &  5 & 14.203  & --0.30  & --0.02     &    0.01     \\
      &  6 & 17.630  & --0.28  &   0.02     &    0.03     \\
      &  7 & 21.240  & --0.32  & --0.02     &  --0.02     \\
      &  8 & 25.000  & --0.32  & --0.02     &  --0.02     \\ \hline
      &  0 & 3.9309  & --0.06  &   0.20     &    0.001     \\
      &  1 & 14.059  &   0.01  &   0.13     &  --0.03     \\
      &  2 & 27.550  & --0.37  & --0.27     &  --0.53     \\
      &  3 & 43.010  & --0.37  & --0.34     &  --0.62     \\
200   &  4 & 60.030  & --0.33  & --0.35     &  --0.64     \\
      &  5 & 78.400  & --0.35  & --0.02     &  --0.01     \\
      &  6 & 97.900  & --0.33  &  $10^{-4}$ &  --0.01     \\
      &  7 & 118.40  & --0.30  &   0.04     &    0.02     \\
      &  8 & 139.90  & --0.32  &   0.01     &  --0.01     \\ \hline
      &  0 & 18.137  & --0.09  &   0.20     & $-10^{-4}$ \\
      &  1 & 64.987  & --0.001 &   0.13     &  --0.03     \\
      &  2 & 127.51  & --0.39  & --0.28     &  --0.54     \\
      &  3 & 199.20  & --0.39  & --0.37     &  --0.65     \\
20000 &  4 & 278.10  & --0.34  & --0.37     &  --0.65     \\
      &  5 & 363.20  & --0.33  & --0.01     &    0.01     \\
      &  6 & 454.00  & --0.40  & --0.06     &  --0.08     \\
      &  7 & 548.90  & --0.32  &   0.02     &  --0.001     \\
      &  8 & 648.50  & --0.31  &   0.01     &  --0.003     \\ \hline
\end{tabular}

\end{center}

\end{document}